\documentclass[aps,prl,twocolumn,10pt]{revtex4-1}

\usepackage{graphicx}
\usepackage{amsmath}
\usepackage{amssymb}

\newcommand{\di}{\mathrm{d}}

\providecommand{\abs}[1]{\left\lvert#1\right\rvert}

\usepackage[all]{xy}
\begin{document}
\title{Spontaneous generation of singularities in paraxial optical fields}
\author{Andrea Aiello$^{1}$}
\email{andrea.aiello@mpl.mpg.de}
\affiliation{$^1$Max Planck Institute for the Science of Light, G$\ddot{u}$nther-Scharowsky-Strasse 1/Bau24, 91058 Erlangen, Germany}
\date{\today}

\begin{abstract}
In nonrelativistic quantum mechanics the spontaneous generation of singularities in smooth and finite wave functions, is a well understood phenomenon also occurring for free particles. We use the familiar analogy between the two-dimensional Schr\"{o}dinger equation and the optical paraxial wave equation to define a new class of square-integrable paraxial optical fields which develop a spatial singularity in the focal point of a weakly-focusing thin lens. These fields are characterized by a single real parameter whose value determines the nature of the
singularity. This novel field enhancement mechanism may stimulate fruitful researches for diverse technological and scientific applications.
\end{abstract}

\maketitle

It has been known in classical optics for some time that there are light beams which tend to abruptly focus while propagating in free space (see, for example, Refs. \cite{Efremidis:10,Chremmos:11,Papazoglou:11,Zhang:11,Ring:12}). In this work we report on similar situations where a singularity suddenly appears in free-propagating  paraxial optical fields. Specifically, we study a family of  paraxial fields, square integrable and almost everywhere smooth, which spontaneously generates a singularity along the propagation axis in the focal point of a thin lens.  One parameter is used to characterize this family  and its value determines the nature of the singularity. According to such value,  the amplitude of the field in the focal point may be either infinite or manifest a cusp singularity.

In his  book on quantum mechanics \cite{PeresBook}, Asher Peres presents an example of a one-dimensional  square-integrable Schr\"{o}dinger wave function, continuous at $t=0$, which evolves into a singular  function at a later time $t>0$. The Schr\"{o}dinger equation for a free particle with mass $m$ in two spatial dimensions
\begin{align}\label{Schro}
\frac{\partial^2 \Psi}{\partial x^2} +  \frac{\partial^2 \Psi}{\partial y^2}  = -i \frac{2 m}{\hbar} \frac{\partial \Psi}{\partial t},
\end{align}
and the paraxial equation for a monochromatic wave of wavelength $\lambda$ and wavenumber $k = 2 \pi /\lambda$ propagating in vacuum
\begin{align}\label{eq10}
\frac{\partial^2 \Psi}{\partial x^2} +  \frac{\partial^2 \Psi}{\partial y^2} = -i 2 k \frac{\partial \Psi}{\partial z},
\end{align}
have the same mathematical form. This analogy is well known and fully documented (see, for example, Ref. \cite{Dragoman} for a comprehensive review). We exploit such analogy to extend Peres' one-dimensional result to paraxial optics, in the spirit of Ref.  \cite{Aiello13}. We find that the situation described by Peres corresponds to a collimated optical beam weakly focalized by a thin lens, and that the time at which the singularity of the Schr\"{o}dinger wave function occurs translates into a propagation distance equal to the focal length of the lens.
%
\begin{figure}[!ht]
\centerline{\includegraphics[scale=3,clip=false,width=.85\columnwidth,trim = 0 0 0 0]{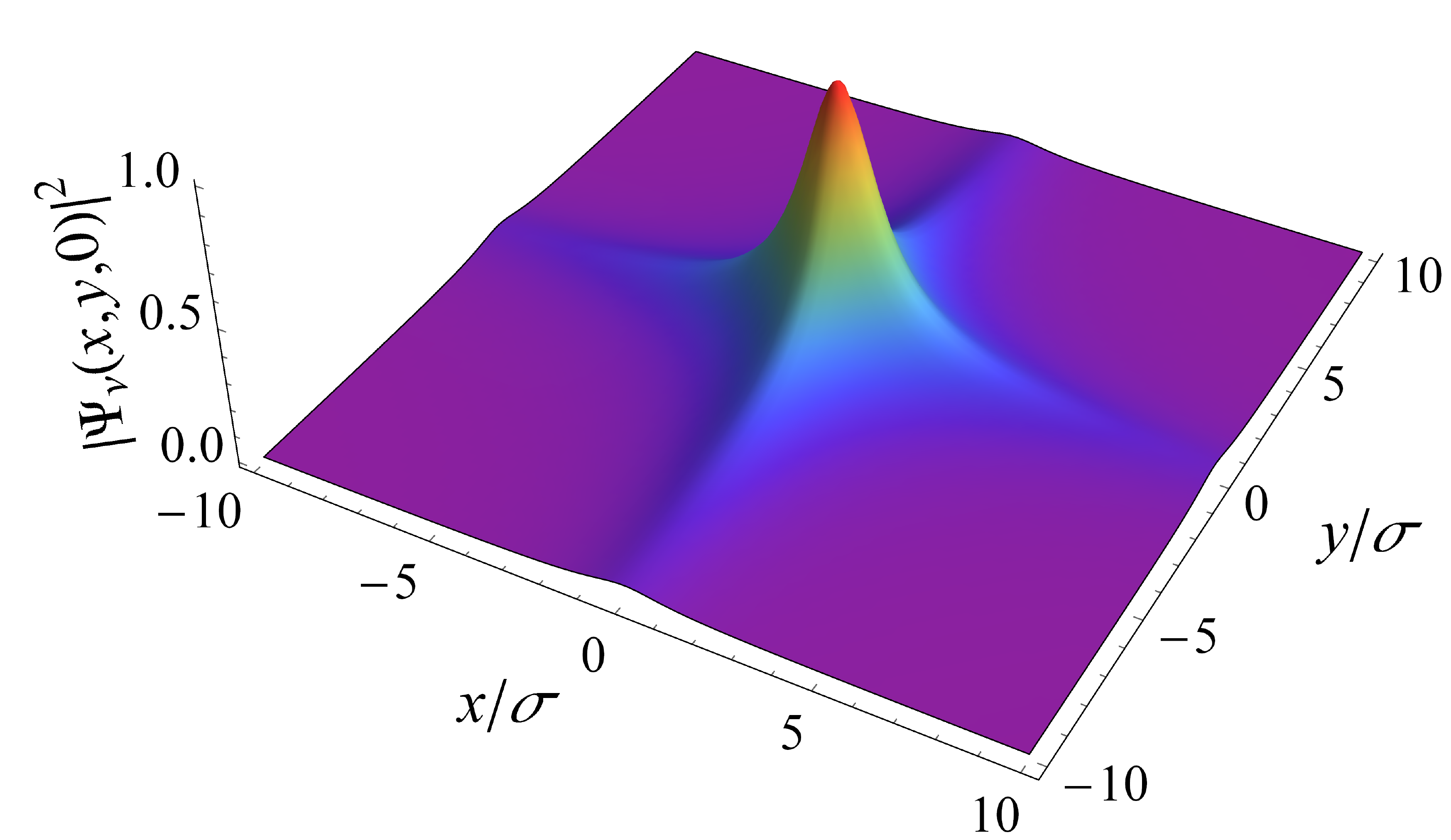}}
\caption{\label{fig1} Plot of the function $\abs{\Psi_\nu(x,y,0)}^2$  for $\nu = 1/3$.}
\end{figure}
%

To begin with, consider a $2$D scalar field with distribution $\Psi(x,y,0)$ on the plane $z=0$. The solution of   \eqref{eq10} for $z>0$ is
\begin{align}\label{eq20}
\Psi(x,y,z) = \frac{k}{2 \pi i z} \int e^{i \frac{k}{2z} \left[(x-\xi)^2 + (y - \eta)^2 \right]} \Psi(\xi, \eta, 0) \di \xi \di \eta,
\end{align}
where the integral is understood over the real $\xi \eta$ plane. The function which multiplies the field $\Psi(\xi, \eta, 0)$ in the integrand of   \eqref{eq20}, is known in classical optics as the Fresnel propagator \cite{GoodmanBook}.
Now, we concentrate on the function $\Psi_\nu(x,y,0) \equiv \psi_\nu(x,0)\psi_\nu(y,0)$, where $\psi_\nu(x,0) = \exp[-i k x^2/(2 f) ]\phi_\nu(x)$, with
\begin{align}\label{eq30}
\phi_\nu(x) = \frac{1}{ \left(1+x^2/\sigma^2 \right)^{\nu}},
\end{align}
where  $f>0$, $\sigma>0$ and $\nu >0$ are given positive constants. Here and hereafter $\phi(x)$ stands for a $1$D field at $z=0$ \emph{before} the lens, while we use $\psi(x,z)$ to denote the same field at $z \geq 0$ \emph{after} the lens, with $\psi(x,0) = \exp[-i k x^2/(2 f) ]\phi(x)$.
$\nu$ is a real-valued parameter that we can vary to obtain different kinds of singularities within the same family of fields specified by \eqref{eq30}. The quantity $\sigma$ determines the spot size of the beam and $f$ is the focal length of a thin lens \cite{GoodmanBook} placed at $z=0$ along the propagation axis $z$.
The function $\Psi_\nu(x,y,0)$ is square integrable  providing that $\nu > 1/4$.
When $\nu=1$,  it represents, up to a phase factor, a so-called Lorentz beam \cite{Gawhary06}.
The cross-shaped intensity profile of $\Psi_\nu(x,y,0)$ is shown in Fig. \ref{fig1}.

\begin{figure}[!ht]
\centerline{\includegraphics[scale=3,clip=false,width=.85\columnwidth,trim = 0 0 0 -50]{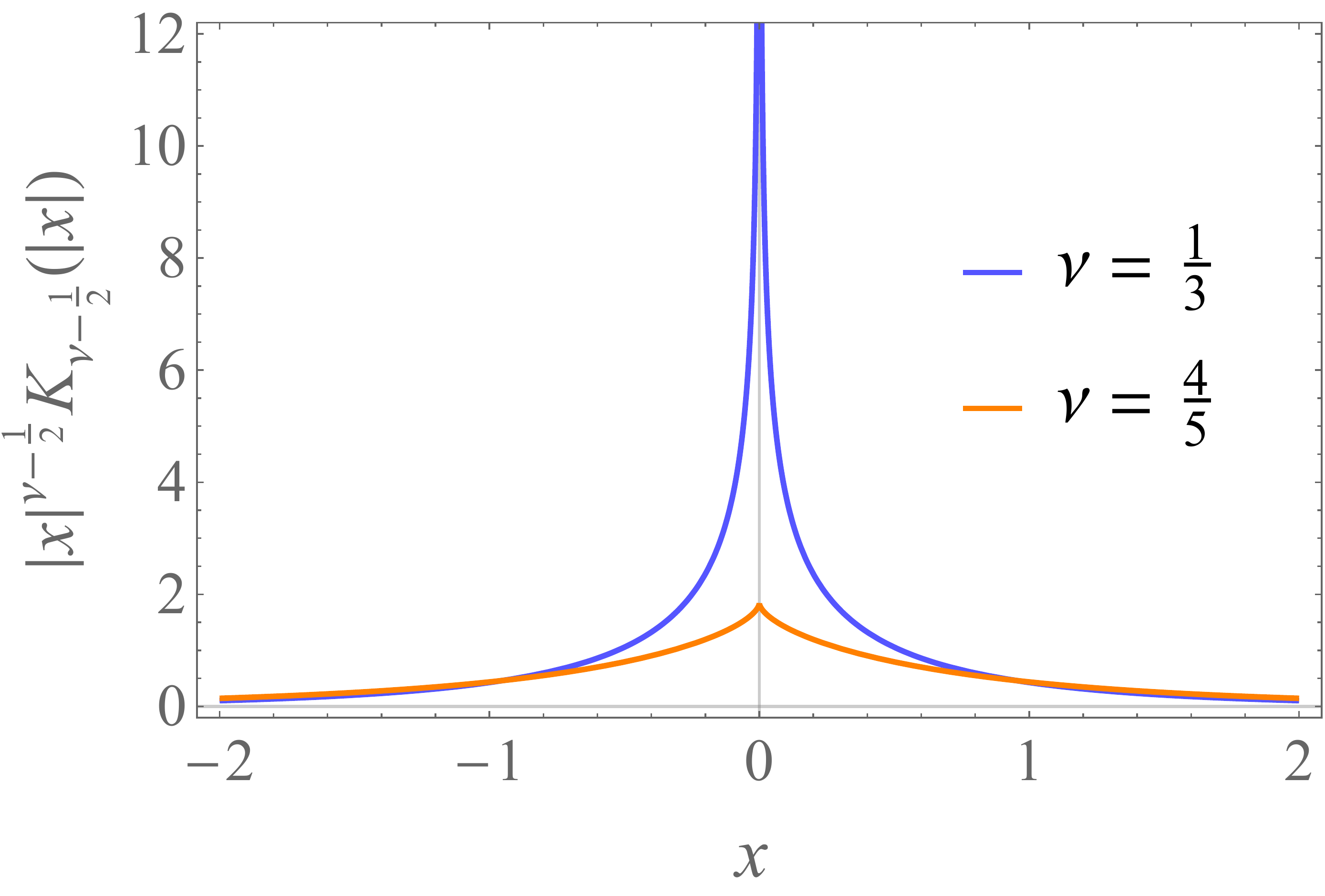}}
\caption{\label{fig2} Plot of the  function $\abs{x}^{ \nu -\frac{1}{2}}\operatorname{K}_{ \nu -\frac{1}{2}}(\abs{x})$ from   \eqref{eq100}, for $\nu = 4/5$ (orange line) and $\nu = 1/3$ (blue line). In the second case the function becomes infinite at $x=0$. }
\end{figure}

Substituting   \eqref{eq30} into   \eqref{eq20}, we obtain
\begin{align}\label{eq70}
\Psi_\nu(x,y,z) =  \frac{k}{2 \pi i z} \,\exp \left[i \frac{k}{2 z} ( x^2 + y^2 )\right] a_\nu (x,z) a_\nu (y,z),
\end{align}
where we have defined the auxiliary function
\begin{align}\label{eq80}
a_\nu (x,z) =  \int\limits_{-\infty}^\infty \frac{\exp\left\{i \frac{k}{2z} \left[-2 x  s +  \left( 1- \frac{z}{f} \right)  s^2 \right]\right\}}{\left(1+{s^2}/{\sigma^2} \right)^{\nu}} \, \di s.
\end{align}
For $x = 0$ and $z = f$  (namely, in the focal point of the lens), the argument of the exponential function above drops to zero and the integrand falls of  as $\abs{s}^{-2 \nu }$ when $\abs{s}$ increases. This implies that  $a_\nu (0,f)$ is infinite for $\nu < 1/2$ and equal to $a_\nu (0,f) = \pi^{1/2} \sigma \Gamma(\nu -1/2)/\Gamma(\nu)$ for $\nu > 1/2$, where $\Gamma (x)$ denotes the gamma function \cite{GR}. Therefore, for $1/4 < \nu < 1/2$ the field $\Psi_\nu$ is square integrable and, nevertheless, infinite at $x=0=y$ and $z=f$.
To explain why this singularity develops from the smooth field $\Psi_\nu(x,y,0)$, we first note that
for $x \neq 0$ and $z \neq f$ the rapid oscillations of the complex exponent make the integrand in   \eqref{eq80} integrable. More precisely, for $1/4 < \nu < 1/2$, the integrand converges for $z \neq f$, it converges conditionally for $z = f$ and $x \neq 0$, and diverges for $z = f$ and $x = 0$.
The rapid oscillations are caused by the mismatch between the radius of curvature (equal to $z$) of the diverging wavefront of the  propagating field and the radius of curvature (equal to $f$) of the converging wavefront generated by the lens. However, at the focal plane $z$ equals $f$, the two radii become identical and the rapid oscillations vanish \cite{Leuchs}.
Then, the remaining integral can be calculated analytically for $\nu> 0$ and the result is
\begin{align}\label{eq100}
a_\nu (x,f) = \frac{\sigma \sqrt{\pi}}{2^{ \nu-\frac{3}{2}}\Gamma(\nu)} \left(
k \sigma \frac{\abs{x}}{f} \right)^{\nu -\frac{1}{2}}  \operatorname{K}_{ \nu -\frac{1}{2}}\left(k \sigma \frac{\abs{x}}{f} \right),
\end{align}
where $\operatorname{K}_{\nu -\frac{1}{2}}$ is the modified Bessel function of the third kind, which is square integrable although it becomes infinite at $x=0$. However, the product $\abs{x}^{ \nu -\frac{1}{2}}\operatorname{K}_{ \nu -\frac{1}{2}}(\abs{x})$ may be either finite or infinite at $x=0$ according to the value of $\nu$. Specifically, for small $\abs{x}$
\begin{align}\label{eq110}
\abs{x}^{ \nu -\frac{1}{2}}\operatorname{K}_{ \nu -\frac{1}{2}}(\abs{x}) \approx  \frac{\Gamma( \nu -1/2)}{2^{3/2-\nu}}
+ \frac{1}{\abs{x}^{1-2 \nu }} \,\frac{\Gamma(- \nu + 1/2)}{2^{ \nu +1/2}} .
\end{align}
This expression  clearly diverges  for small $\abs{x}$ and $\nu \neq 1/2 + m$, (with $m$ positive integer).
Figure \ref{fig2} shows the function $\abs{x}^{ \nu -\frac{1}{2}}\operatorname{K}_{ \nu -\frac{1}{2}}(\abs{x})$ for $\nu = 4/5$ and $\nu = 1/3$. The  value  $\nu = 4/5$ only generates a cusp singularity at $x=0$, while for $\nu = 1/3$ the function becomes  infinite at $x=0$.
%
%
\begin{figure}[!hb]
\centerline{\includegraphics[scale=3,clip=false,width=.9\columnwidth,trim = 0 0 0 40]{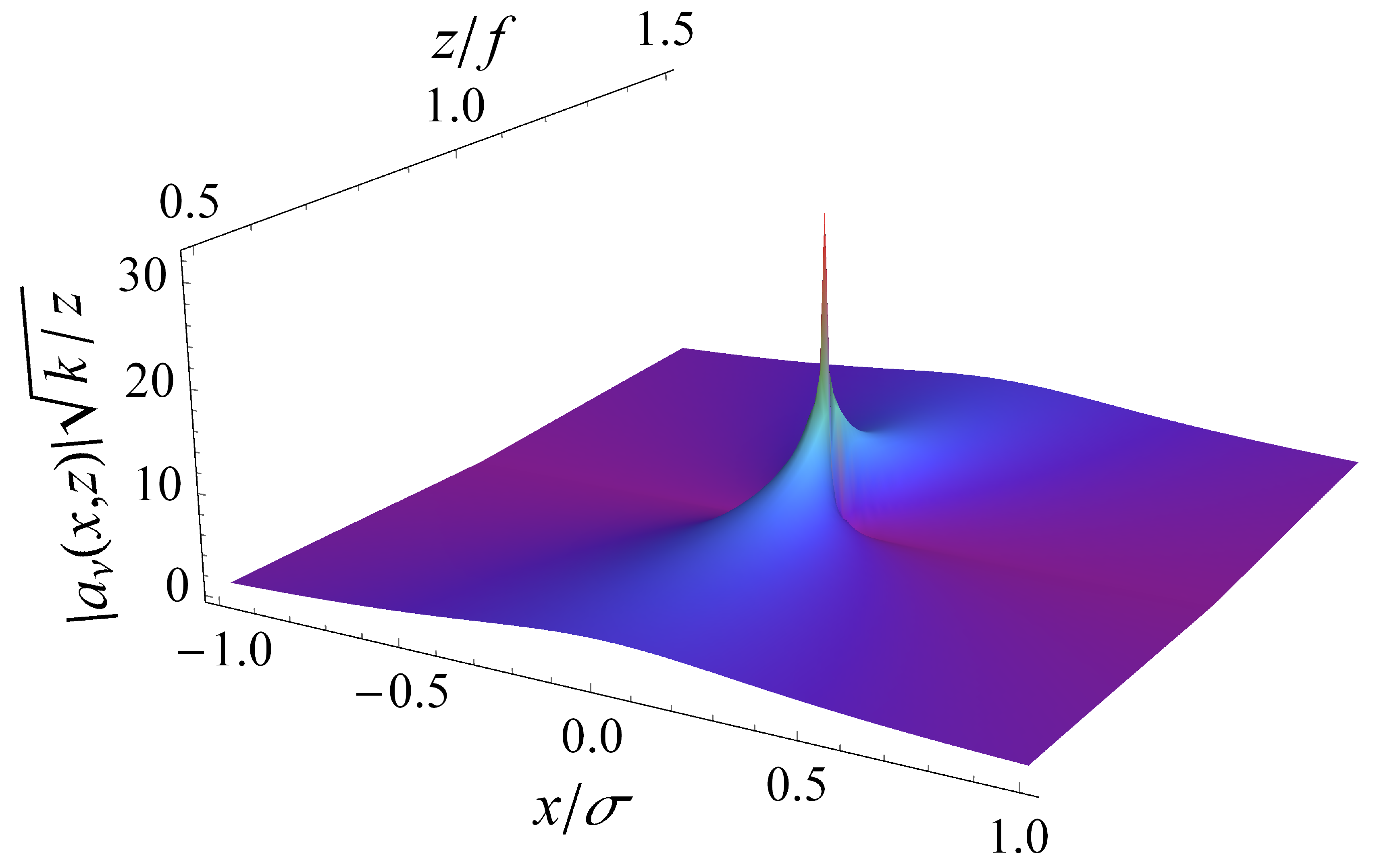}}
\caption{\label{fig3} Plot of the dimensionless function $\abs{a_\nu (x,z)}  \sqrt{k/z}$  for $\nu = 4/5$, $f = 250 \, \text{mm}$ and $\lambda = 500 \, \text{nm}$, as resulting from a numerical integration. The function is steep but finite at $x=0$.}
\end{figure}
%

For $z\neq f$ the integral in   \eqref{eq80} can be calculated either exactly via numerical integration (see Fig. \ref{fig3}) or approximately by the method of stationary phase \cite{MandelBook}. For the last case, the exponent is stationary when $s = x/(1-z/f)$
and a standard calculation gives
\begin{align}\label{eq130}
a_\nu (x,z)\approx  \frac{\left({\pm  2 \pi \, i \,z}/{k}\right)^{1/2}}{ \left[ 1 + \frac{x^2/\sigma^2}{ \left( 1 - z/f \right)^2} \right]^{\nu}}  \frac{\exp \left(-i \frac{k}{2z} \frac{x^2}{1-z/f} \right) }{\abs{1-z/f}^{1/2}} ,
\end{align}
where the upper or lower sign is taken, according as $z\lessgtr f$. This approximation  breaks down in the very neighborhood of $z = f$ and $x \neq 0$ where the stationary point ceases to be a critical point of the first kind and tends to the endpoints $s = \pm \infty$. However, we could verify the validity of    \eqref{eq130} outside this critical region, by comparison with numerical integration.
%
\begin{figure*}[!ht]
\centerline{\includegraphics[scale=3,clip=false,width=1.7\columnwidth,trim = 0 0 0 0]{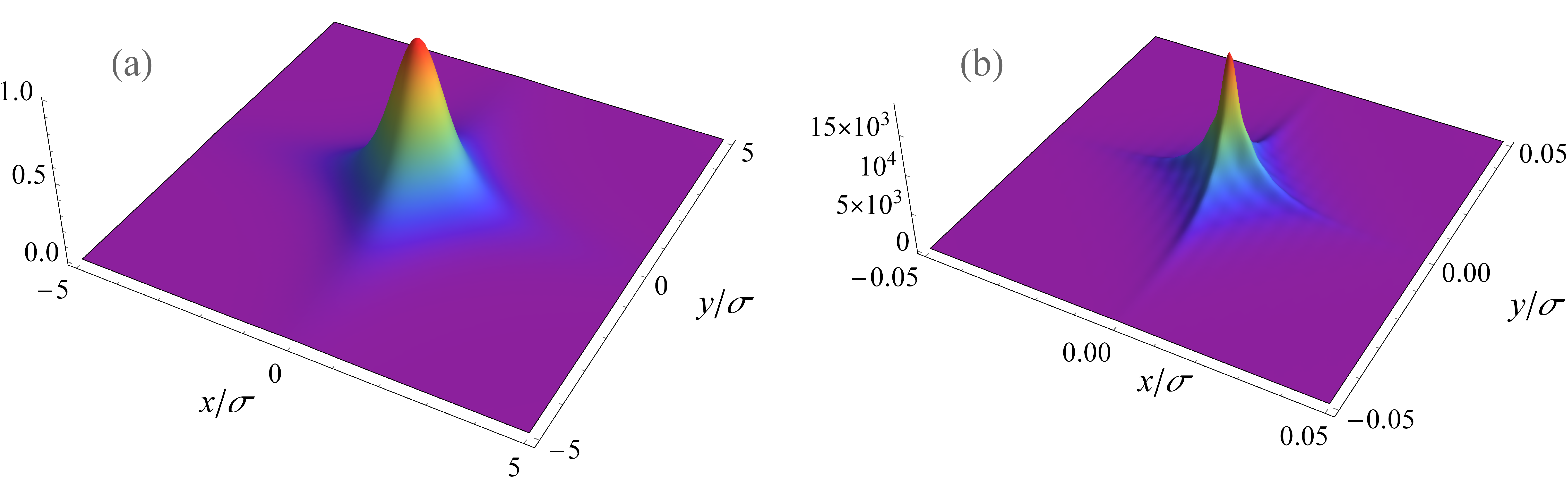}}
\caption{\label{fig4} Intensity profiles $I_\nu =\abs{\Psi_\nu (x,y,z)}^2/\abs{\Psi_\nu (0,0,0)}^2$ at (a) the origin $z=0$ and (b) in  the focal plane $z/f = 1$ of the lens, calculated with the values $\nu = 4/5$, $\sigma = 1 \, \text{mm}$, $f = 250 \, \text{mm}, \,\lambda = 500 \, \text{nm}$ and $L= 30 \, \text{mm}$. In this case $L/f =0.12$ and the paraxial approximation  is still valid. Note the different scales of the $x$-$y$ axes for the two plots. }
\end{figure*}
%

The first moments $\langle x \rangle_{\Psi_\nu} $ and $ \langle y \rangle_{\Psi_\nu} $ of the distribution $\abs{ \Psi_\nu(x, y, z)}^2$ are finite and equal to zero whenever  $\nu > 1/2$. The second moment $\langle x^2 + y^2 \rangle_{\Psi_\nu} $  exists providing that $\nu > 3/4$ and gives a measure  of the beam waist. Using some results from \cite{PhysRevE.59.7152}, it is not difficult to show that
\begin{align}\label{eq140}
\frac{\langle x^2 + y^2 \rangle_{\Psi_\nu}}{\sigma^2} =  \frac{1/2}{\nu - 3/4} \left( 1 - \frac{z}{f}\right)^2  + \frac{z^2}{z_0^2}  \frac{\nu (4 \nu -1)}{4(2 \nu + 1)},
\end{align}
where $\nu > 3/4$ and we have defined $z_0 \equiv k \sigma^2/2$, which fixes a unit of longitudinal length equivalent to the Rayleigh range of a Gaussian beam \cite{MandelBook}. As expected, since both terms in   \eqref{eq140} are non-negative for $\nu > 3/4$, the minimum value of $\langle x^2 + y^2 \rangle_{\Psi_\nu}$ is achieved in the focal plane $z=f$, where
\begin{align}\label{eq150}
\left.\sqrt{ \langle x^2 + y^2 \rangle_{\Psi_\nu}}\right|_{z=f} =\frac{f \sigma}{z_0}  \sqrt{\frac{\nu (4 \nu -1)}{4(2 \nu + 1)}}.
\end{align}
The square-root term on the right side of this equation  varies between $0$ and $1/2$ for $1/4\leq \nu \leq 1$. Therefore, in order to have a small spot size is desirable to have ${f \sigma}/{z_0} = f \lambda/(\pi \sigma)\ll \lambda$. This cannot be achieved in the paraxial regime of propagation where $\sigma/f \equiv \tan \theta \approx \theta \ll 1$ and ${f \sigma}/{z_0} =  \lambda/(\pi \tan \theta) \gg \lambda$. That is, sub-wavelength focusing is not possible for $\nu > 3/4$.

Any physically realizable optical system has necessarily a finite aperture. This means that for a real-world beam Eq. (6) should be replaced by
\begin{align}\label{new10}
A_\nu (x,z;L) \equiv  \int\limits_{-L}^L\frac{\exp \left\{i \frac{k}{2z} \left[-2 x  s +  \left( 1- \frac{z}{f} \right)  s^2 \right]\right\}}{\left(1+{s^2}/{\sigma^2} \right)^{\nu}} \, \di s,
\end{align}
where $A_\nu (x,z;\infty) = a_\nu (x,z)$ and $2L$ is the diameter of the lens.  The asymptotic expansion
of $ A_\nu (0,f;L) = 2 L \,{}_2F_1(1/2,\nu;3/2;-L^2/\sigma^2)$, where  $ \/_2F_1(\alpha,\beta;\gamma;z)$ denotes the hypergeometric function \cite{GR}, can be calculated via a Taylor expansion around $L = \infty$, which gives
\begin{align}\label{new20}
 A_\nu (0,f;L) \approx\frac{\sigma\pi^{1/2} \Gamma(\nu -1/2)}{\Gamma(\nu)} + \frac{2 \sigma}{1-2 \nu}\left( \frac{L}{\sigma} \right)^{1-2\nu}.
\end{align}
This equation shows that  the singularity of $\Psi_\nu(0,0,f)$ is dissolved in the passage to a finite aperture because $A_\nu (0,f;L) \approx  L^{1-2\nu}$, which is always finite for $L<\infty$ and becomes infinite for $L \to \infty$ when $\nu<1/2$. This means that the origin of the singularity at $x=0$ and $z=f$ is rooted in the infinite lateral extent of the field $\phi_\nu$. However, this mechanism is very different from the one generating a
  Dirac delta singularity in the focal plane of a lens of infinite aperture illuminated by a plane wave. In this case the singularity arises at the cost of the infinite amount of energy carried by the incident field. Conversely, $\phi_\nu$  is square integrable for $\nu>1/4$.
In passing, we note that the apparent singularity in \eqref{new20} for $\nu = 1/2$ is an artifact of the expansion, as it can be seen by calculating  explicitly $ A_{1/2} (0,f;L) = 2 \, \sigma \operatorname{arcsinh}(L/\sigma)$. A comparison between the intensity distribution of the field before the lens and in the focal plane of the latter, is shown in Fig. \ref{fig4} for $\nu = 4/5$ and  $L= 30 \, \text{mm}$.

For the realistic case of finite $L$, it is interesting to compare the focusing efficiency of the ``singular'' field $\phi_\nu(x)$ with respect to a Gaussian field $\phi_G(x) = \exp(-x^2/\sigma_G^2)$ and a ``rectangular'' field $\phi_R(x) =  \operatorname{rect} (x/\sigma_R)$, where $\operatorname{rect} (x)$ denotes the rectangular function \cite{GoodmanBook}.
From $\phi_\nu(0)=\phi_G(0)=\phi_R(0)=1$ it follows that the three fields  have the same peak intensity at $z=0$. The parameters $\sigma_G$ and $\sigma_R$ which set the waists of the Gaussian and rectangular fields, respectively, are fixed by imposing that these fields carry the same energy of the singular field, namely
\begin{align}\label{new30}
\int\limits_{-L}^L \abs{\phi_\nu(x)}^2\, \di x = \int\limits_{-L}^L \abs{\phi_G(x)}^2\, \di x= \int\limits_{-L}^L \abs{\phi_R(x)}^2\, \di x.
\end{align}
 This equation implies that both $\sigma_G$ and $\sigma_R$ depend on  the width $\sigma$ of the singular field and on the aperture diameter $2L$. For example, using \eqref{new30} with $\sigma = 1 \, \text{mm}$ and $L= 30 \, \text{mm}$, we find $\sigma_G = 4.27 \, \text{mm}$ and $\sigma_R =5.36 \, \text{mm}$. Figure \ref{fig5} below shows the behavior of the three fields $\phi_\nu, \phi_G$ and $\phi_R$ at $z=0$ (before the lens) and at $z=f$ (in the focal plane of the lens). Ceteris paribus, the singular field $\phi_\nu$ exhibits a focal intensity distribution  narrower and higher  than standard Gaussian and rectangular fields.
%
\begin{figure}[!hb]
\centerline{\includegraphics[scale=3,clip=false,width=.95\columnwidth,trim = 0 0 0 0]{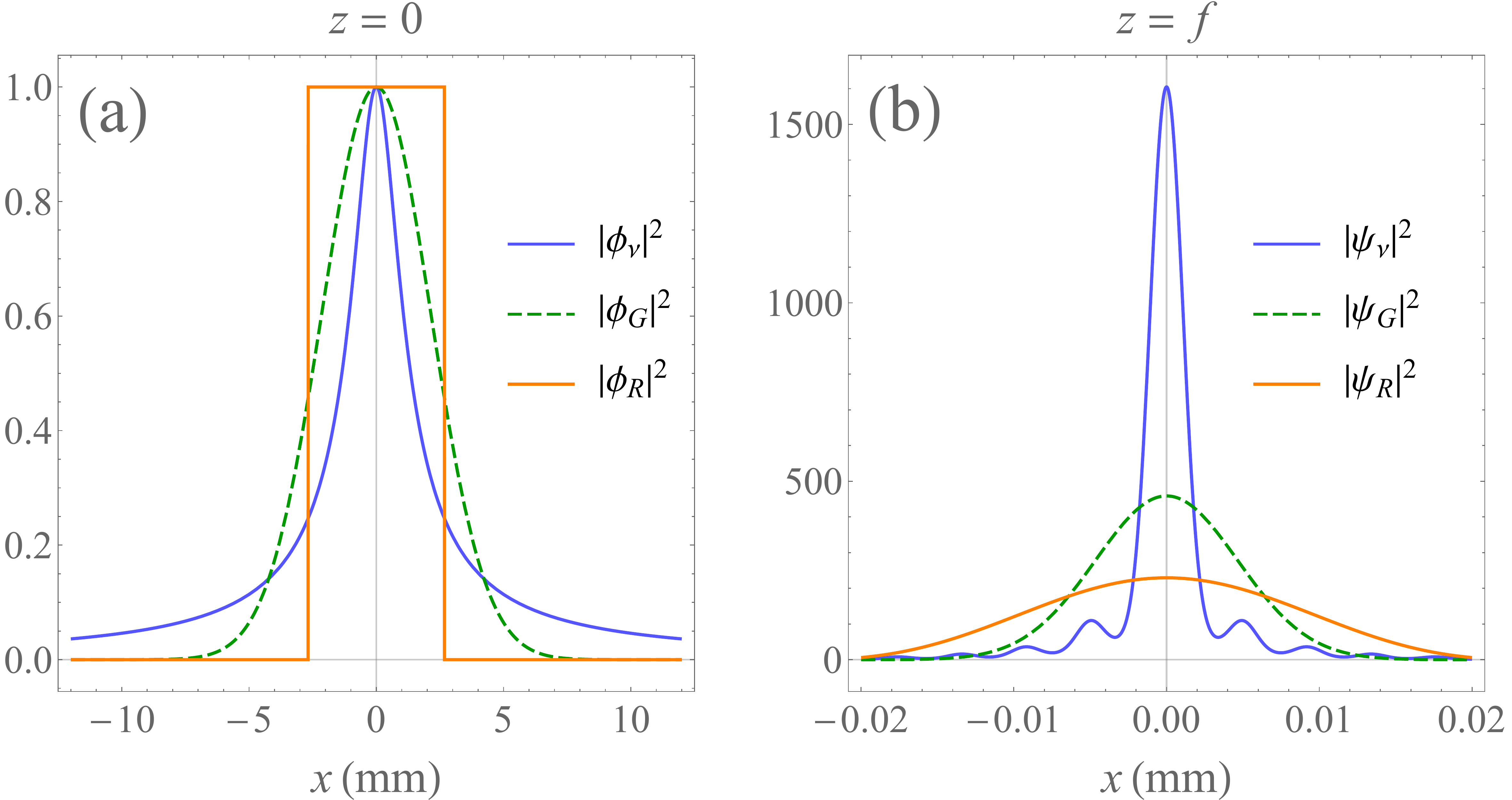}}
\caption{\label{fig5} (a) Intensity distributions of the  fields $\phi_\nu$ (blue line), $\phi_G$ (dashed green line) and $\phi_R$ (orange line) at $z=0$.  (b) Intensity distributions of the same fields as in (a), but evaluated on the focal plane $z=f$. In both plots the width of the Gaussian and rectangular function has been chosen according to \eqref{new30}, giving $\sigma_G = 4.27 \, \text{mm}$ and $\sigma_R =5.36 \, \text{mm}$ for $\nu = 1/3$, $\sigma = 1 \, \text{mm}$ and $L= 30 \, \text{mm}$. The focal length $f$ of the lens is fixed to $f = 250 \, \text{mm}$. For the paraxial Gaussian beam considered here, the equation $z=f$ sets the focal plane to a good approximation.}
\end{figure}
%

The results illustrated in Fig. \ref{fig5} (b) may induce to think about a possible violation of the diffraction limit by the singular field $\phi_\nu$. However, this is not the case, as clearly illustrated in Fig. \ref{fig6} below, where the singular field $\phi_\nu$ is compared to a plane wave illuminating uniformly the aperture and carrying the same amount of total energy. In fact, the diffraction limit is determined by the intensity distribution generated by the uniform illumination which is, according to Fig. \ref{fig6} (b), less wide than the distribution of  $\abs{\psi_\nu}^2$.
%
\begin{figure}[!ht]
\centerline{\includegraphics[scale=3,clip=false,width=.95\columnwidth,trim = 0 0 0 0]{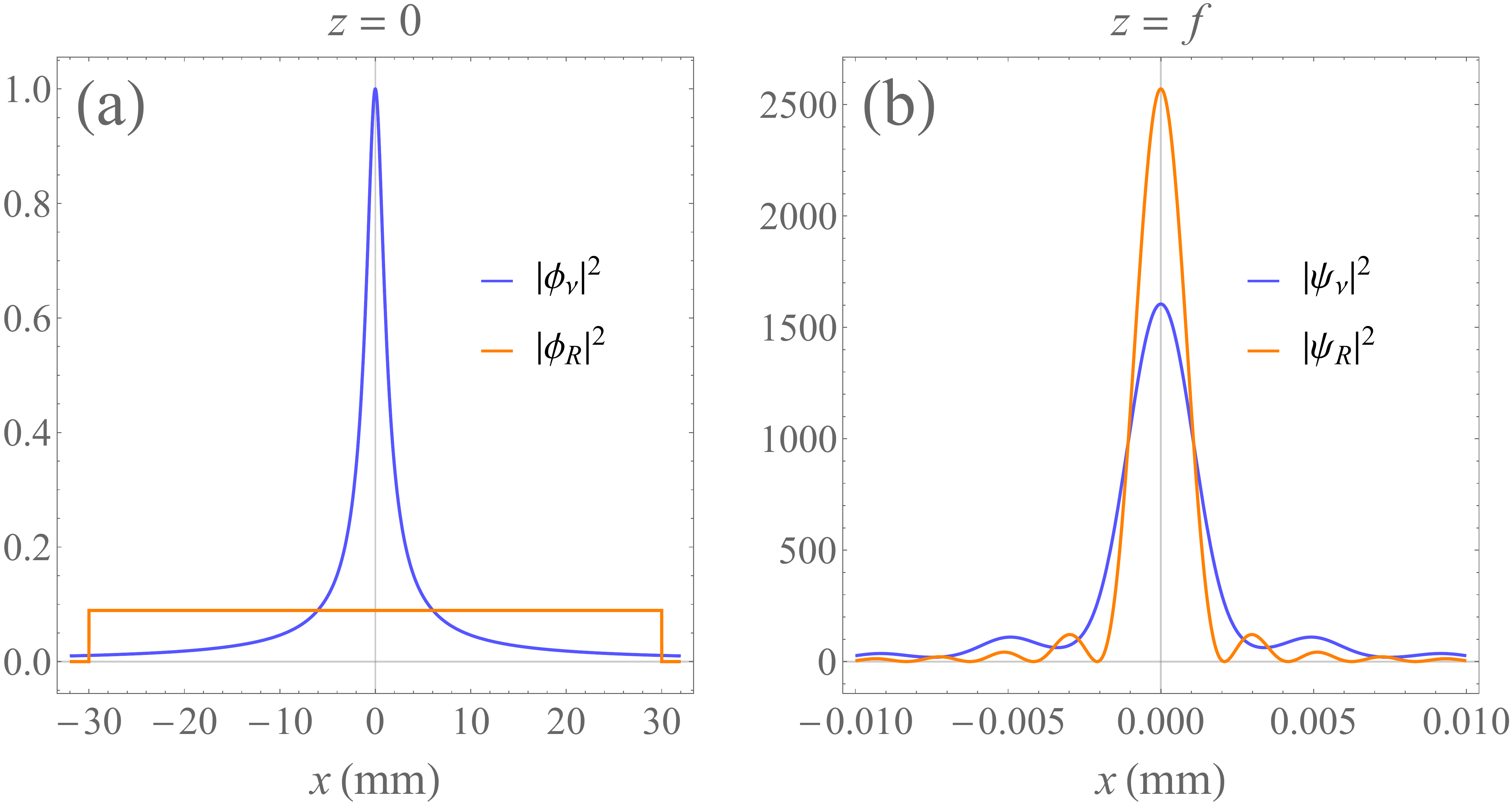}}
\caption{\label{fig6} (a) Intensity distributions at $z=0$ of the singular field $\phi_\nu$ (blue line) and the re-normalized rectangular field $Z_R^{1/2}\phi_R$ (orange line), where $Z_R = \/_2F_1(1/2,2 \nu;3/2;-1/4)$ with  $\nu = 1/3$ and $\sigma_R = 2 L$.  (b) Intensity distributions of the same fields as in (a), but evaluated at $z=f$. Here $ \/_2F_1(\alpha,\beta;\gamma;z)$ denotes the hypergeometric function \cite{GR} and $\sigma = 1 \, \text{mm}$, $L= 30 \, \text{mm}$,  $f = 250 \, \text{mm}$.}
\end{figure}
%

To summarize, we have presented and studied a family of paraxial optical beams which spontaneously develop a singularity while propagating in free space. The use  of either antennas or nonlinear optical devices is not required to observe this effect. However, an optical system not limited by a finite aperture is required to achieve the singularity.
Perhaps the most valuable characteristics of these beams is that they are represented by square integrable functions. 
Because of this, they do not seem to belong to the wide class of Airy beams \cite{Airy} and X-waves  amply studied in recent years (see, for example, \cite{BerryB,Lekner09,PhysRevLett.108.163901} and \cite{PhysRevLett.90.170406,PhysRevE.69.036608}).
Due to the finiteness of  our beams, their singularities do not require a nonphysical infinite amount of energy to manifest. Nevertheless, the local amplitude of the field at a singular point may grow unboundedly.
This promising field enhancement mechanism may foster further interesting researches in classical and quantum nonlinear optics.

We thank Gerd Leuchs for suggesting the study of focusing properties of free fields and for fruitful discussions.
We also thank Norbert Lindlein for providing numerical simulations and for insightful comments on the manuscript.


\begin{thebibliography}{10}
\bibitem{Airy}
For the special values $\nu=5/6$ and $\nu=7/6$ the modified Bessel function in \eqref{eq100} is linked to the Airy function via the equations $\operatorname{K}_{\pm 1/3}(\zeta) = \pi \sqrt{3/z} \operatorname{Ai}(z)$ and $\operatorname{K}_{\pm 2/3}(\zeta) = -\pi \sqrt{3}/z \operatorname{Ai}'(z)$, respectively, where $z = (3 \zeta/2)^{2/3}$.
\end{thebibliography}
\end{document}